\documentclass{article}
\usepackage{graphics}
\title{Slow crack growth: models and experiments}
\author{S. Santucci, L. Vanel, S. Ciliberto \\
Laboratoire
de Physique, CNRS UMR 5672, \\
Ecole Normale Sup\'erieure de Lyon \\
46 all\'ee d'Italie, 69364 Lyon Cedex 07, France}

\begin{document}
\maketitle
\abstract{The properties of slow crack growth in brittle materials
are analyzed both theoretically and experimentally. We propose a
model based on a thermally activated rupture process. Considering a
2D spring network submitted to an external load and to thermal
noise, we show that a preexisting crack in the network may slowly
grow because of stress fluctuations. An analytical solution is found
for the evolution of the crack length as a function of time, the
time to rupture and the statistics of the crack jumps. These
theoretical predictions are verified by studying experimentally the
subcritical growth of a single crack in thin sheets of paper. A good
agreement between the theoretical predictions and the experimental
results is found. In particular, our model suggests that the
statistical stress fluctuations trigger rupture events at a
nanometric scale corresponding to the diameter of cellulose
microfibrils.
} 
\section{Introduction}
\label{intro}

Research on fracture has received a lot of attention from the
physics community. This interest is obviously motivated by the
numerous practical benefits that a better understanding of the
fracturing processes in solid materials would bring to many
engineering domains. But also from a theoretical point of view, the
study of damaging processes in heterogeneous materials appears
crucial in different fields of physics, and brings forward many
challenging questions in particular in statistical physics
\cite{herrmann90,Alava}.

Here, we are interested in slow rupture processes observed when a
material is submitted to a constant load below a critical rupture
threshold (creep test). It is well known that the delay time (or
lifetime) of the material before complete macroscopic rupture
strongly depends on the applied stress. Thermodynamics has slowly
emerged as a possible framework to describe delayed rupture of
materials since early experiments have shown temperature dependence
of lifetime with an Arrhenius law \cite{Brenner,Zhurkov}. The
current understanding is that subcritical rupture can be thermally
activated with an activation energy which depends on the applied
stress. For elastic materials, several statistical models have been
recently proposed in order to predict the lifetime
\cite{Golubovic,Pomeau1,Buchel,Kitamura97,Roux,Scorretti} and the
average dynamics of a slowly growing crack \cite{Santucci1}. In
these models, it is assumed that the thermal noise inside the
material induce stress fluctuations that will nucleate small cracks
if the stress becomes larger than the local rupture threshold of the
material. These models are interesting because in certain conditions
they allow the prediction of the lifetime of a sample as a function
of the macroscopic applied stress. However the  test of this idea is
not simple because other models based on viscoelastic retardation of
the crack formation \cite{Schapery86,Langer,Chudnovsky} may in some
cases explain the formation of the delayed crack.

The purpose of this paper is to review a series of experiments and
theoretical studies that have been performed in order to test in
some details the activation models in brittle materials,  that is
for materials whose stress-strain curve remains elastic till
failure. Since the model we have chosen is a two dimensional one,
the experiments are also performed in a situation very close to a 2D
geometry. Specifically we have studied the slow propagation of a
single crack in a thin sheet of paper which in dried atmosphere is a
brittle material. As we will see, this geometry allows an accurate
comparison with the theoretical predictions.

The paper is organized as follows: in section 2, we review the main
properties of the model and we summarize the main predictions; in
section 3, we describe the experimental apparatus and the properties
of the averaged crack growth; in section 4, the statistics of the
crack jumps is discussed. We conclude in section 5.

\section{A model for the slow crack growth}\label{section_model}

It is very well known that when a material is submitted to a
constant stress (creep test) it breaks after a certain time $\tau$
which is a function of the applied stress. We are interested in
modeling this phenomenon for brittle materials. Specifically, we
want to derive the dependence of the lifetime $\tau$ on the applied
stress and of the damage, i.e. the number of broken bonds, as a
function of time. A common approach to describe the creep rupture of
materials is to introduce a time-dependent creep compliance or a
rate of rupture which is a power law of the applied stress
\cite{Paris}. Instead of assuming a phenomenological law for the
creep behavior, we use a statistical approach which takes into
account the fact that at equilibrium there are always statistical
fluctuations of stress with a variance which depends on the actual
temperature of the material. These local stress fluctuations may be
larger than the local rupture stress of the material, thus producing
local damage.

\subsection{A thermally activated crack nucleation}

The starting point is the Griffith theory for fracture in a brittle
material\cite{Griffith,Lawn}. Griffith's prediction of a critical
crack size beyond which there is rupture, i.e. irreversible and fast
crack growth, is derived from a potential energy taking into account
the elastic energy due to the applied stress $\sigma$ and the
surface energy $\gamma$ needed to open a crack as a function of a
unique order parameter, the crack length $L$ \footnote{see
\cite{Buchel} for a generalization of this approach taking into
account crack opening.}. For a bidimensional geometry consisting of
a flat sheet with a crack perpendicular to the direction of stress,
the potential energy per unit thickness of the sheet reads:
\begin{equation}\label{Griffith}
 E_G(L) =  - \frac{{\pi L ^2 \sigma ^2 }}{{4Y}} + 2\gamma L + E_0
\end{equation}
where $Y$ is the Young modulus and $E_0$ is the elastic energy in
the absence of crack.
\begin{figure}[hbt!]
\centerline{
\resizebox{0.5\columnwidth}{!}{\includegraphics{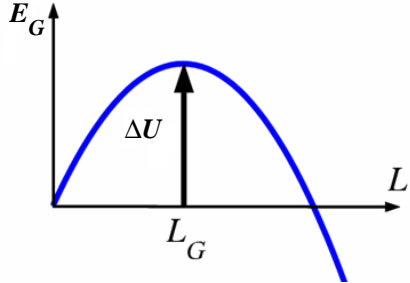}}}
    \caption{Sketch of the Griffith potential energy $E_G$
    as a function of crack length $L$.}
    \label{Grif}
\end{figure}
A typical example of the Griffith potential as a function of $L$ is
shown on fig.\ref{Grif}. This potential energy reaches a maximum for
the Griffith length $L_G=4Y\gamma / (\pi \  \sigma^2)$ which in this
case coincide with a critical length $L_c$ beyond which no stable
state exists except the separation of the solid in two broken
pieces. Thus, a stressed solid without a crack appears to be in a
metastable state as long as no crack with a critical length
nucleates\cite{Golubovic,Pomeau1}.

Several authors \cite{Brenner,Zhurkov,Golubovic,Pomeau1,Buchel} have
used models essentially inspired by Griffith's energy concept and
considered that the nucleation of a crack with critcal size could be
thermally activated. Then, the lifetime should follow an Arrhenius
law:
\begin{equation}\label{eqtau}
  \tau  \sim \exp \left( \frac{\Delta U}{k_B T}
  \right)
\end{equation}
where $k_B$ is Boltzmann constant and  $T$ temperature. The energy
barrier for the two-dimensional case can be obtained from
eq.\ref{Griffith} and scales as $\Delta U=E_G(L_G)\sim\sigma^{-2}$.
Note that for a three-dimensional geometry the potential energy
would give a barrier $\Delta U=E_G(L_G)\sim\sigma^{-4}$.

If this approach has permitted to reproduce and interpret
qualitatively some experimental results\cite{Pauchard,Ciliberto},
quantitatively the temperature fluctuations appear too weak to be
able to overcome the Griffith's energy barrier $\Delta U= E_G(L_G)$,
except in the case of a Griffith length having atomic scale.
Furthermore, there is another problem since this choice of energy
barrier implicitly assumes that there is a possibility for the crack
to explore reversible states of crack length between the initial and
the critical one. If such a process could occur when the Griffith
length is at atomic scale, this is certainly not true if one
consider the experimental case where a preexisting macroscopic crack
is growing progressively. In fact, this approach cannot describe any
kind of dynamics where the rupture process appears to be
irreversible.

\subsection{An irreversible and thermally activated crack
growth}\label{meangrowth}

In order to overcome the problem of irreversibility in thermally
activated crack growth, let us start from a different point of view
where irreversible rupture events can be caused by stress
fluctuations due to thermal noise.

The uniaxial loading state of an homogeneous solid at fixed
temperature is described by its free energy density:
$\varphi(\sigma) = \sigma_m^2/2 Y$, where $\sigma_m$ is the internal
stress of the material. Treating stress as a fluctuating internal
variable in a fixed volume $V$, the probability to find a given
stress is proportional to a Boltzmann factor $\exp(-\varphi
V/k_BT)$. Expanding free energy about the equilibrium position
$\sigma_m$, the distribution of stress $\sigma_f$ is :
\begin{equation}\label{distribution}
  g(\sigma_f) \simeq \frac{1}
  {\sqrt {2\pi \langle \Delta \sigma\rangle^2}}
  \exp \left[ -\frac{\left(\sigma_f-\sigma_m\right)^2}{2\langle \Delta \sigma\rangle^2} \right]
\end{equation}
where  $\langle \Delta \sigma\rangle^2 = k_BT/(V\partial^2
\varphi/\partial \sigma^2)=k_BT Y/V$\cite{Diu}. When a crack is
present, stress concentration increases the probability that
breaking occurs at the crack tip rather than anywhere else. We
assume that stress distribution at the crack tip remains the same as
eq.~(\ref{distribution}) despite the strong divergence of stress and
the breakdown of linear elasticity. Since the stress intensity
factor $K \approx \sigma \sqrt{L}$ gives a measure of stress
intensity close to the crack tip for a crack with length $L$ when
the external load is $\sigma$, we choose to work directly with $K$
and the stress at the crack tip is $\sigma_m=K/\sqrt{\lambda}$ where
$\lambda$ is a microscopic characteristic scale. Here we assume that
the material is mainly elastic but  at the scale $\lambda$  it
becomes discontinuous. For example in a perfect crystal, the only
such scale would be the atomic scale, in a fibrous materials, like
paper or fiber glass, we have an intermediate mesoscopic scale, i.e.
the typical fiber size, and in the 2D elastic spring network the
size of the elementary cell of the network. Within this description
the threshold for rupture at the crack tip will be given by a
critical value of stress intensity factor $K_c$ as is usual
laboratory practice.

In order to model crack rupture as a thermally activated process, we
assume that a volume $V$ of the material will break if the
fluctuating stress $K$ in this volume becomes larger than the
threshold $K_c$. The breaking probability of the volume element is
then:
\begin{equation}\label{Proba}
P(K>K_c)=\int_{K_c}^{\infty} g(K) \lambda^{-1/2} dK
=\int_{U_c}^\infty \frac{e^{-U_f}dU_f}{\sqrt{\pi U_f}}
\end{equation}
where
\begin{equation}\label{speed1a}
U_f(K_f)={(K_f-K)^2 V \over 2Y \ \lambda \ k_B T},\ \ \ \ \
U_c=U_f(K_c) \ \ \ \ \ \mathrm{and} \ \ \ \ \ K_f=\sigma_f
\sqrt{\lambda}.
\end{equation}
The lifetime of this volume element is \cite{Roux}:
$\tau_V=-\tau_e/\ln(1-P)$ where $\tau_e$ is an elementary time
scale (typically, an inverse vibrational frequency). Then, the
velocity $v$ of the crack tip is simply $v=\lambda/\tau_V$. As
long as $U_c\gg 1$ (in other words, when the energy barrier is
larger than $k_B T$), we have $P\ll 1$, thus $\tau_V\simeq
\tau_e/P$, $v\simeq\lambda P /\tau_e$ and we can approximate the
integral in eq.(\ref{Proba}) to get:
\begin{equation}\label{motion}
 v=\frac{\  dL }{\  d t}
  \simeq {\lambda \over \tau_e} \sqrt {\frac{Y\ \lambda \ k_BT}{2\pi V}}
  \frac{1}{K_c  - K}\exp \left[ - \frac{\left(K_c  - K\right)^2 V}{2\ Y \lambda \ k_BT}
  \right].
\end{equation}
Because $K$ is a function of crack length $L$, eq.(\ref{motion}) is
in fact a differential equation for the crack evolution. To solve
this equation requires additional approximations since the
dependence of stress intensity factor on crack length is non-linear.
In numerical simulations of 2D networks of springs with thermal
noise and in experiments one observes that all the relevant dynamics
of the crack growth occurs for $L\simeq L_i$ and $L<L_c$. Then, the
stress intensity factor can be written:
\begin{equation}
  K \approx \sigma \sqrt{L} = \sigma \sqrt{L_i+(L-L_i)}
  \simeq K_i \left[1+\frac{1}{2}\left(L-L_i\right)\right]
\end{equation}
where the last equality is a reasonable approximation giving less
than a $2\%$ error on stress intensity factor as long as $L <3/2
L_i$.  Another approximation will be to take $K=K_i$ in the
pre-factor of the exponential, because neglecting the variation in
stress intensity factor leads only to a logarithmic correction of
the crack velocity. As a consequence of the last approximation, the
crack velocity will tend to be underestimated.

Solution of the differential equation (\ref{motion}) is then :
\begin{equation}\label{eqgrowth}
  t = \tau \left[1 - \exp \left( -\frac{L-L_i}{\zeta}\right)\right]
\end{equation}
where $\tau$ gives the lifetime of the sample before fast rupture:
\begin{equation}\label{eqlifetime}
  \tau  = \tau_0\exp \left[ \frac{(K_c  - K_i )^2 V}{2Y\lambda \ k_BT}
  \right]  \qquad \rm{with} \qquad \tau_0=\frac{\tau_e}{\lambda} \ \frac{ 2\ L_i }{K_i } \   \sqrt{\frac{2\  \pi \ Y\ \lambda \  k_BT}{
  V}}
\end{equation}
and $\zeta$ is  a characteristic growth length:
\begin{equation}\label{eqzeta}
  \zeta  = \frac{2 \ Y \  \lambda \ k_BT \ L_i}{V\ K_i(K_c-K_i)}
\end{equation}

Note that the crack velocity : $\frac{\  dL}{\  d t} =
\zeta/(\tau-t)$, diverges as time comes closer to lifetime $\tau$,
which simply means that when time $\tau$ is reached slow crack
growth due to thermal activation is no longer the driving mechanism,
and a crossover towards fast dynamic crack propagation will occur.
The lifetime $\tau$ appearing in eq.~(\ref{eqlifetime}) follows an
Arrhenius law with an energy barrier
\begin{equation}\label{eqDU}
  \Delta U= \left[ \frac{(K_c  - K_i )^2 V}{2Y\lambda} \right]
\end{equation}
which is a  function of initial and critical stress intensity
factors. A similar scaling for the energy barrier was found by
Marder \cite{Marder}.

\subsection{Description of an intermittent dynamics and crack pinning}\label{jump}

\subsubsection{A modified Griffith energy barrier due to lattice
trapping effect}

The time evolution of the crack length predicted by
eq.(\ref{eqgrowth}) has to be considered an average one. In reality
both numerical simulations and experiments (see section
\ref{section_experiment}) show that the crack tip progresses by
jumps of various size and it can spend a lot of time in a fixed
position. This dynamics can be understood by considering the
existence of the characteristic microscopic scale $\lambda$
introduced in the previous section. Indeed, the elastic description
of a material at a discrete level leads to a lattice trapping effect
\cite{Thomson86} with an energy barrier which has been estimated
analytically \cite{Marder}. The other important effect of the
discreteness is that $L_c$ becomes larger than $L_G$.
\begin{figure}[hbt!]
\centerline{
\resizebox{0.4\columnwidth}{!}{\includegraphics{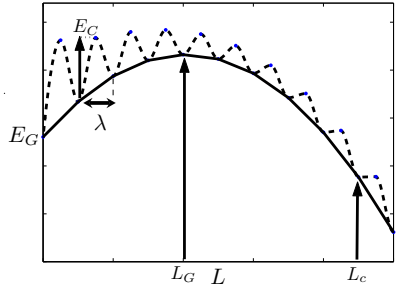}}}
    \caption{Sketch of the Griffith potential energy $E_G$ as a function of crack length $L$ with constant applied stress (solid line).
    The energy barriers $E_C$ and the discretization scale $\lambda$ are represented by the dashed curve.}
    \label{dino}
\end{figure}
To get a physical picture of the trapping we may  consider a 2D
square lattice of linear springs where the crack corresponds to a
given number $n$ of adjacent broken springs as described in
\cite{Santucci1}. The lattice is loaded with a constant stress
$\sigma$ and we estimate the minimum increase in potential energy
needed to bring at the breaking threshold the spring at the tip of
a crack of length $n\lambda$. This spring is submitted to a stress
$\sigma_m=K/\sqrt{\lambda}=\sqrt{n} \ \sigma$.  It is clear that
in order to move the crack tip from the position $n$ to $n+1$ the
stress $\sigma_f$ on the crack tip has to reach at least the
material stress threshold for rupture $\sigma_c$. Thus to estimate
the energy barrier $E_c$ that the system has to overcome
 to move the crack from $n$ to $n+1$ we consider again the free energy density defined in
  sec.~\ref{meangrowth}. The
increase $E_f$ of the free energy density produced by a an
increase of the stress $\sigma_f$ on the crack tip can be computed
by a Taylor expansion of the bulk elastic free energy density
around the equilibrium value of the stress $\sigma_m$, i.e.
$E_f(\sigma_m)\simeq(\sigma_f-\sigma_m)^2/2Y$. As the crack moves
only if $\sigma_f\ge \sigma_c$, the energy barrier that the system
has to overcome is $E_c(\sigma_m)\simeq (\sigma_c-\sigma_m)^2/2Y$.
To each position of the crack tip corresponds a different value of
the energy barrier since $\sigma_m$,the stress  at the tip,
increases with the crack length. Once the spring breaks, the crack
moves by at least one lattice spacing $\lambda$. The equilibrium
potential energy of the whole system is given by the Griffith
energy eq.(\ref{Griffith}). These simple arguments have been
checked in a numerical simulation of the 2D spring lattice loaded
with a constant stress. The energy barrier $E_c$ is obtained by
applying an external force on the spring at the crack tip and
computing the change in elastic energy of the whole lattice and
the work done by the constant force at the boundaries. The results
of the simulation is plotted in Fig.\ref{dino} where we represent
the energy barrier of trapping schematically\footnote{Note that
the trapping barrier exists in fact for a fixed discrete length of
the crack. The exact energy path that joins one equilibrium length
to the next one is unknown} (dashed line) and the Griffith energy
(continuous line). In agreement with previous analysis
\cite{Marder}, we find that the crack length $L_c$ at which the
energy barrier becomes zero is much larger than  the Griffith
length $L_G$ where the equilibrium potential energy reaches its
maximal value.

\subsubsection{Irreversible thermally activated stepwise growth}

We now recompute the mean crack speed $v$ by considering the
thermally activated and irreversible motion of a crack in the rugged
potential energy landscape introduced above. Below $L_c$, the energy
barriers $E_c(\sigma_m)$ trap the crack in a metastable state for an
average time $\tau_p$ depending on the barrier height. Irreversible
crack growth is a very reasonable assumption when $L>L_G$ since the
decrease in equilibrium potential energy makes more likely for the
crack to open than to close. As already mentioned in the previous
section,  when a fluctuation $\sigma_f$ occurs, it will increase
locally the free energy per unit volume by
$E_f(\sigma_m)\simeq(\sigma_f-\sigma_m)^2/2Y$.  The energy $E_f$ can
be used by the crack to overcome the barrier. If there are no
dissipative mechanisms the crack will grow indefinitely when $L>L_G$
as the barriers get smaller and smaller and the release of elastic
energy helps to reach a more energetically favorable position. We
introduce a simple mechanism of crack arrest assuming that after
overcoming the energy barrier the crack looses an energy identical
to the barrier size and does not gain any momentum from the elastic
release of energy (experimentally, dissipation will come from
acoustic wave emissions, viscous or plastic flow, etc.). When the
crack reaches the next trap it still has an energy $E_f-E_c$ which
might be sufficient to overcome the next barrier. For a given
fluctuation energy $E_f$, the crack will typically have enough
energy to overcome a number of barriers
$n=E_f(\sigma_m)/E_c(\sigma_m)$ and make a jump of size $s=n
\lambda$ (the decrease of $E_c(\sigma_m)$ with $\sigma_m$ during a
jump of size $s$ has been neglected). Thus $s$ is related to the
quantities defined in eq.(\ref{speed1a}), that is $s=U_f \ \lambda /
U_c$. The probability distribution for $E_f$ is explored at each
elementary step $\tau_0$, while the probability distribution of step
size is explored after each average time $\tau_p$ spent in the trap.
In order to relate the two probabilities, we express the mean
velocity $v$ in a different way as the ratio of the average step
size to the average trapping time:
\begin{equation}\label{speed2}
v=\frac{\int_\lambda^\infty s p(s) ds}{\tau_p}
\end{equation}
where $p(s)$ is the distribution of the jump amplitudes at a given
$K$. Identifying eq.(\ref{speed2}) with $v=\lambda P /\tau_0$ (see
paragraph \ref{meangrowth} and eq.(\ref{Proba})) and using the above
mentioned hypothesis that $s=U_f \ \lambda / U_c$ we can write that
$sp(s)ds=\exp{(-U_f)} \ (\pi U_f)^{-1/2} \ dU_f$. From the  the
normalization condition of the probability ($\int_{\lambda}^\infty
p(s)ds=1$), we obtain the probability distribution :
\begin{equation}\label{distri}
p(s)=N(U_c)\frac{\sqrt{\lambda}e^{-s/\xi}}{2 s^{3/2}}
\end{equation}
 where $N(U_c)=\left[e^{-U_c}-\sqrt{\pi
U_c}\rm{erfc}(\sqrt{U_c})\right]^{-1}$ and $\xi=\lambda/U_c$. We
find a power law with an exponent $3/2$ and an exponential cut-off
with a characteristic length $\xi\sim(K_c-K)^{-2}$ diverging at the
critical stress $K$. Incidently, we note that this probability has a
form similar to sub-critical point probability distributions in
percolation theory \cite{Stauffer91}. From eq.(\ref{distri}), we can
compute from this distribution the average and variance of step
sizes:

\begin{equation}\label{mean}
\langle s \rangle=N(U_c)\frac{\lambda\sqrt{\pi}}{2
\sqrt{U_c}}\rm{erfc}(\sqrt{U_c})
\end{equation}

\begin{equation}\label{var}
\langle s^2 \rangle=N(U_c)\frac{\lambda^2\sqrt{\pi}}{4
U_c^{3/2}}\left(\rm{erfc}(\sqrt{U_c})+2
\sqrt{\frac{U_c}{\pi}}e^{-U_c}\right)
\end{equation}

We obtain two asymptotical behaviors. When the relative energy
barrier is high ($U_c\gg 1$), $\langle s \rangle\simeq\lambda$ and
$\langle s^2 \rangle\simeq\lambda^2$. In this limit, there is only
one step size possible. When the relative energy barrier becomes low
($U_c\ll 1$), we predict a divergence at critical point : $\langle s
\rangle\sim(K_c-K)^{-1}$ and $\langle s^2 \rangle\sim(K_c-K)^{-3}$.
Then, the crack velocity is expected to be dominated by the critical
divergence of crack jumps.

\section{The slow crack growth in a brittle material}
\label{section_experiment}

The theoretical predictions of the previous section have been
checked in an experiment in a quasi  two dimensional geometry.
Specifically, the samples are thin sheets of paper with an initial
cut which are submitted to a constant applied stress
\cite{Santucci3,Santucci2}.

\subsection{Experimental set-up}
\begin{figure}[hbt!]
\centerline{
{\resizebox{0.5\columnwidth}{!}{\includegraphics{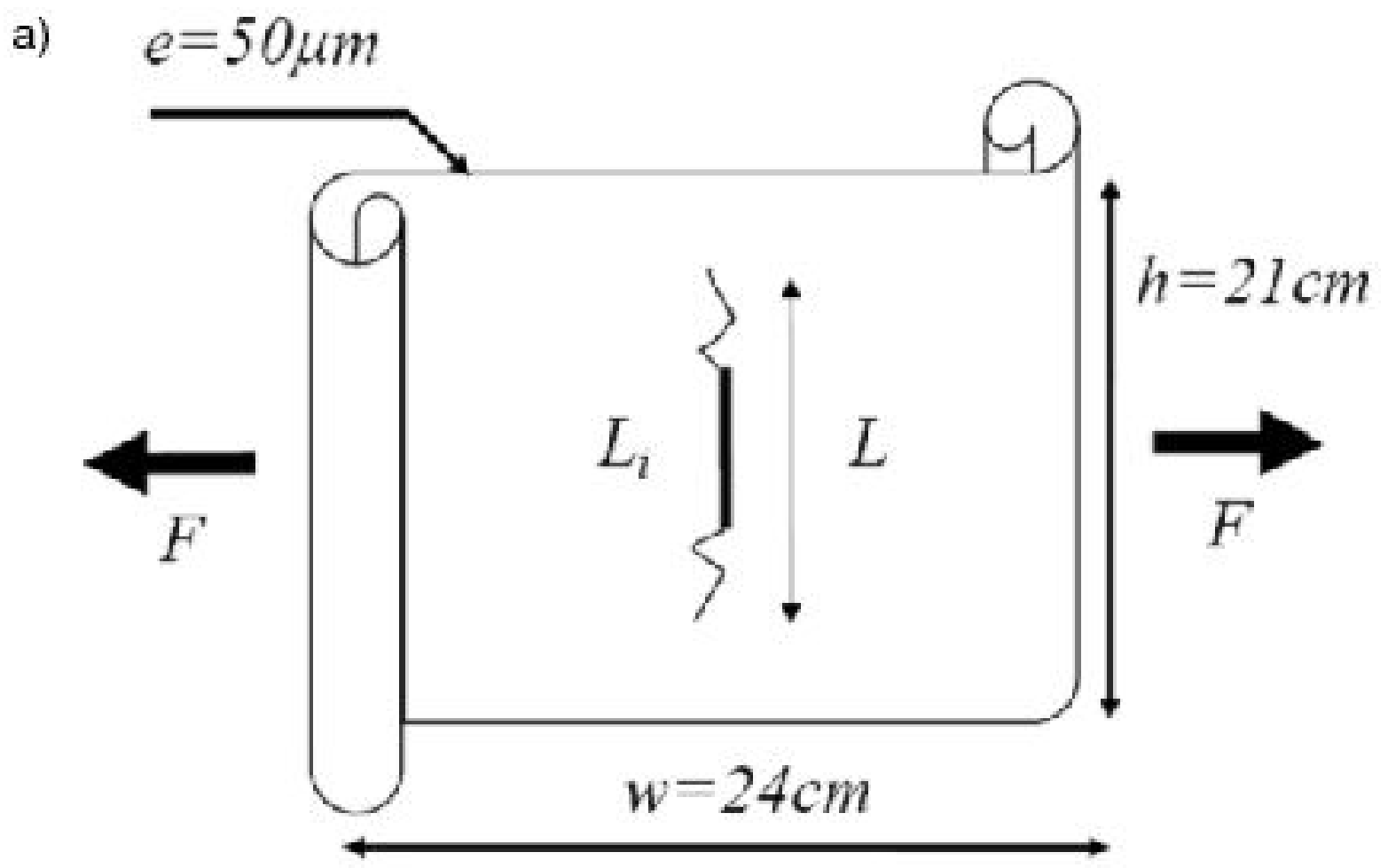}}}
{\resizebox{0.5\columnwidth}{!}{\includegraphics{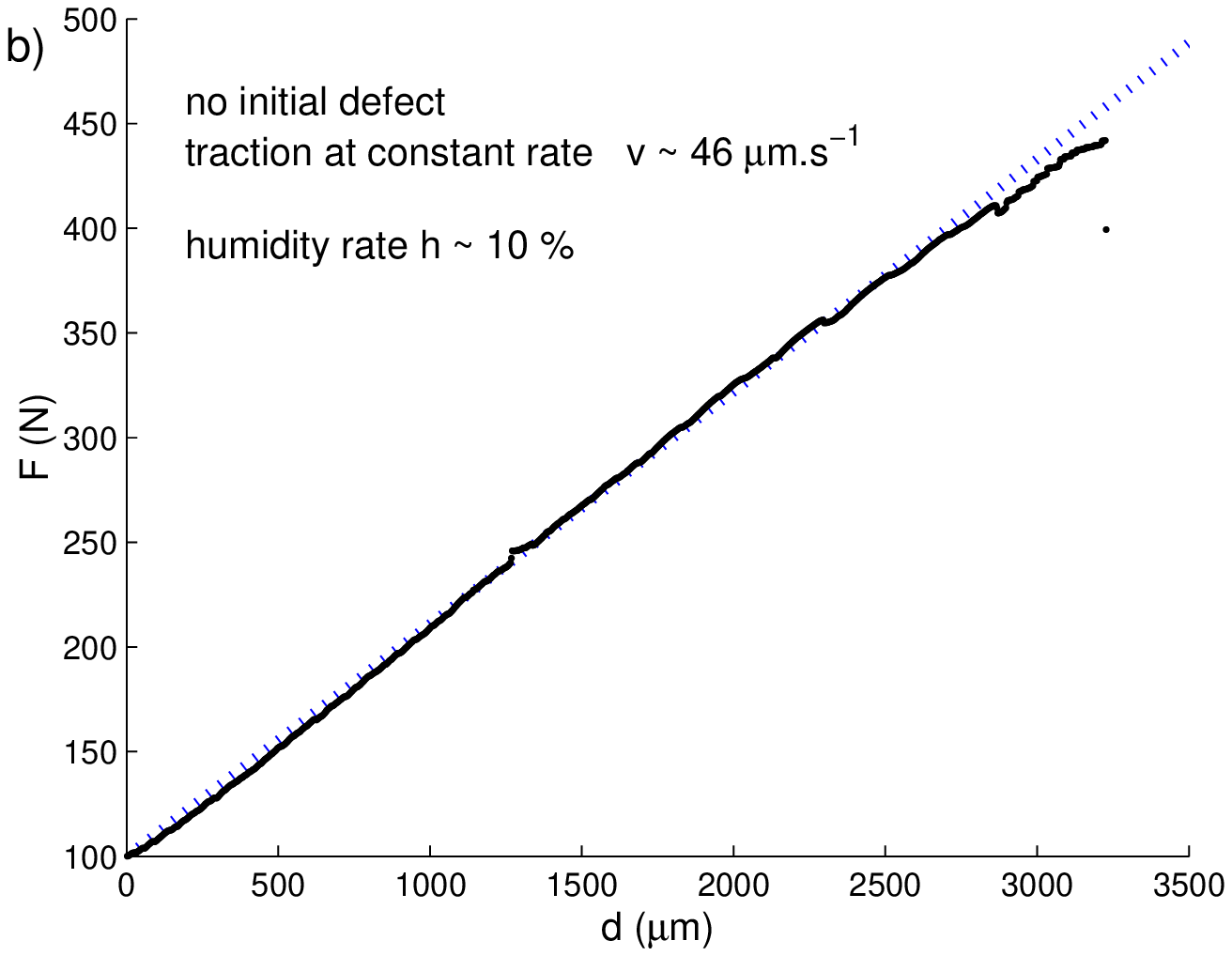}}}
} \caption{ a) Sample geometry. b) Linear dependence between
applied force and elongation until rupture. } \label{f.1}
\end{figure}

Crack growth is obtained by loading in mode 1 at a constant force
$F$ a sheet of fax paper (Alrey) with an initial crack in the center
(fig.~\ref{f.1}a). The sample dimensions are : height $h = 21
\rm{cm}$, width $ w=24\rm{cm}$, and thickness $e = 50 \rm{\mu m}$.

The experimental set-up consists of a tensile machine driven by a
motor (Micro Controle $UE42$) controlled electronically to move step
by step (Micro Controle $ITL09$). The paper sheets are mounted on
the tensile machine with both ends attached with glue tape and
rolled twice over rigid bars clamped on jaws. The motor controls the
displacement of one jaw ($400$ steps per micrometer) while the other
jaw is rigidly fixed to a force gage (Hydrotonics-TC). The tensile
machine is placed in a box with the humidity rate stabilized at
$5\%$. In order to work on samples with the same initial crack shape
and length $L_i$, we use calibrated razor blades mounted on a
micrometric screw and we initiate a macroscopic crack precisely at
the center of the sheet. The samples are loaded by increasing the
distance between the jaws such that the resulting force $F$ is
perpendicular to the initial crack direction. A feedback loop allows
us to adjust the displacement in order to keep the applied force $F$
constant with a precision better than $0.5 \rm{N}$ and a response
time less than $10 \rm{ms}$. From the area $A$ of a cross-section of
the sheet, $A$ being approximatively constant, we calculate the
applied stress $\sigma=F/A$.

\subsubsection{Physical properties of
paper}

Sheets of fax paper in a dry atmosphere break in a brittle manner.
This is evidenced by the elastic stress-strain dependence which is
quasi-linear until rupture (fig.~\ref{f.1}b). Another sign that
rupture is essentially brittle is given by the very good match
between the two opposite lips of the fracture surfaces observed on
post-mortem samples.
\begin{figure}[hbt!]
\centerline{ {\resizebox{0.8\columnwidth}{!}
{\includegraphics{review_fig5.eps}}}} \caption{Scanning electron
microscopy performed at GEMPPM (INSA Lyon) shows the microstructure
of our samples and even the defects on the surface of a single
fiber.} \label{paperstructure}
\end{figure}
A sheet of paper is a complex network of cellulose fibers. Scanning
electron microscopy (see fig.\ref{paperstructure}) on our samples
shows fiber diameters between $4 $ and $50 \rm{\mu m}$ with an
average of $18 \rm{\mu m}$. Cellulose fibers are themselves a bundle
of many microfibrils. Cellulose microfibrils have a cristalline
structure (therefore, they are very brittle) and are consistently
found to have a diameter $d=2.5 nm$ \cite{Jakob}.

The mechanical properties of paper depend crucially on the humidity
rate. To get reproducible results, the fax paper samples are kept at
least one day at a low humidity rate ($< 10 \%$) and during the
experiment ($\simeq 5\%$). At constant humidity rate ($hu \simeq
5\%$) and room temperature, the Young modulus of the fax paper
sheets is typically $Y = 3.3\,10^9 \rm{N.m^{-2}}$.

\subsubsection{Direct observation and image analysis}

We light the samples from the back. A high resolution and high speed
digital camera (Photron Ultima 1024) collects the transmitted light
and allows us to follow the crack growth. We observe that the global
deformation of the paper sheet during a creep experiment is
correlated in a rather reproducible way to the crack growth whatever
the rupture time. We use this property to trigger the camera at
fixed increment of deformation (one micron) rather than at fixed
increment in time. This avoids saturation of the onboard memory card
when the crack growth is slow and makes the acquisition rate faster
when the crack grows faster and starts to have an effect on global
deformation. We acquire $2$ frames at $250 \rm{fps}$ at each trigger
and obtain around one thousand images per experiment.
\begin{figure}[hbt!]
\centerline{
{\resizebox{0.7\columnwidth}{!}{\includegraphics{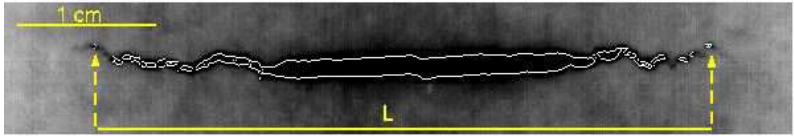}}}}
\caption{ Extraction of the projected crack length $L$ from the
crack contour detected.} \label{extract_L}
\end{figure}
Image analysis is performed to extract the length of the crack
projected on the main direction of propagation, i.e. perpendicular
to the direction of the applied load (fig.~\ref{extract_L}).
Although the crack actually follows a sinuous trajectory, its
projected length $L$ gives the main contribution to the stress
intensity factor $K$ which we compute as: $K= \sigma
\sqrt{\frac{\pi}{2}L \psi(\pi L /2 H)}$, where $\psi(x)=\tan x/x$ is
a correction due to the finite height $H$ of the samples
\cite{Lawn}.

\subsection{Experimental results}

\subsubsection{The single crack growth}
For a given initial crack length $L_i$, subcritical crack growth is
obtained by applying a constant force $F$ so that $K(L_i)$ is
smaller than a critical rupture threshold of the material $K_c$
above which fast crack propagation would occur \cite{Santucci3}.
During an experiment, the crack length increases, and so does the
stress intensity factor $K(L)$. This will cause the crack to
accelerate until it reaches a critical length $L_c$ for which
$K(L_c)=K_c$.

\begin{figure}[hbt!]
\centerline{{\resizebox{0.5\columnwidth}{!}{\includegraphics{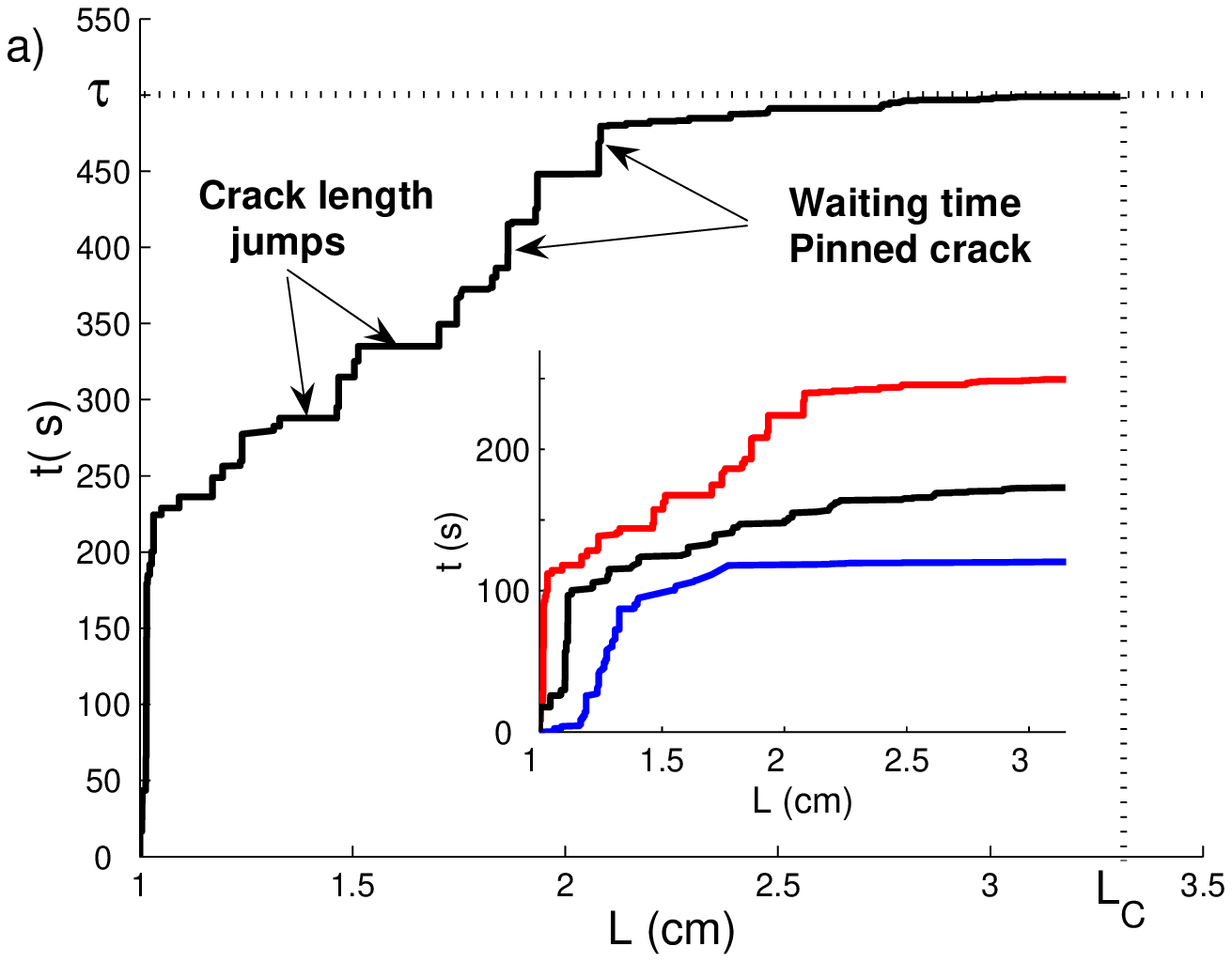}}}
{\resizebox{0.5\columnwidth}{!}{\includegraphics{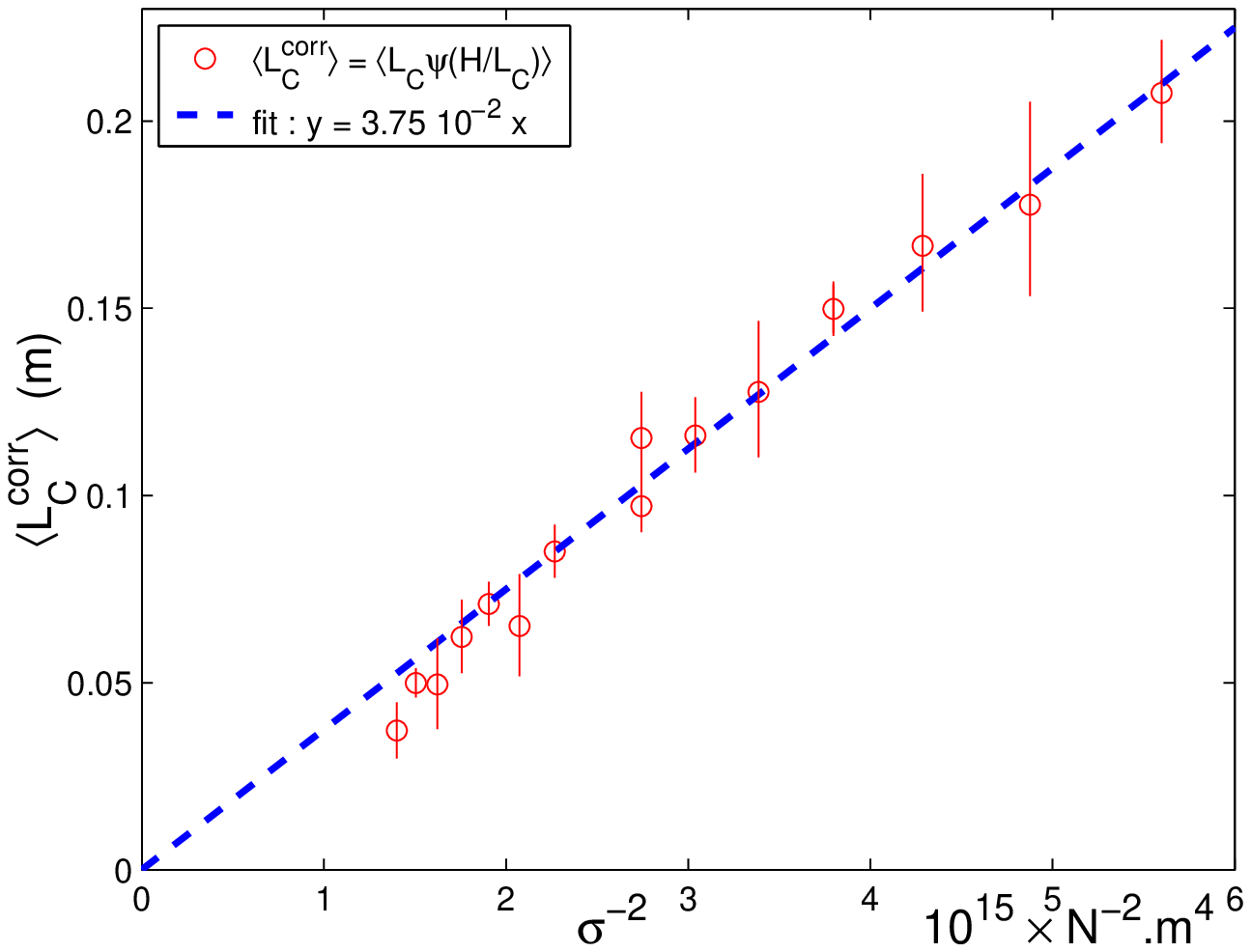}}}}
\caption{ a) Typical stepwise growth curve for a creep experiment
with an initial crack length $L_i=1\rm{cm}$ submitted to a constant
load $F=270\rm{N}$. The lifetime of the sample is $\tau =500\rm{s}$
and the critical length $L_c=3.3\rm{cm}$. In insert, a strong
dispersion is observed in crack growth profile and in lifetime for
$3$ creep experiments realized in the same conditions. b) Critical
length of rupture $L_c$ as function of the inverse square of the
applied stress $1/\sigma^2$. The dashed line represents the best
linear fit $y=K_c^2 x$. Its slope permits us to estimate the
critical stress intensity factor $K_c$. Note that we introduce the
finite height corrections in the critical length $L_c^{corr}$.}
\label{singlegrowth}
\end{figure}

On fig.~\ref{singlegrowth}a) we show a typical growth curve obtained
during a creep experiment with an applied force $F=270\,\rm{N}$ and
an initial crack length $L_i=1\rm{cm}$. Since time to rupture $\tau$
is a statistical quantity, we prefer to plot time evolution as a
function of the crack length. We observe that the crack growth is
actually intermittent. Essentially, there are periods of rest during
which the crack tip is pinned and does not move, and other  moments
when the crack suddenly opens and advances of a certain step size
$s$. The crack advances by jumps until it reaches a critical length
$L_c$ where the paper sheet breaks suddenly. On
fig.~\ref{singlegrowth} b) measurements of $L_c$ are used to
estimate the critical stress intensity factor $K_c=\sigma \sqrt{\pi
L_c^{corr}/2}$, where we include the finite height corrections in
the critical length $L_c^{corr}$. We find $K_c=6 \pm
0.5\,\rm{MPa.m^{1/2}}$.

Beyond $L_c$, the crack runs across the whole sample (about $18
\rm{cm}$ in this case) in less than one second, with a crack speed
$v > 300 \rm{m.s^{-1}}$. For the same experimental conditions (same
stress, same initial crack length, same temperature and same
humidity rate), we observe a strong dispersion in growth curves and
in lifetime while the critical length seems to be rather well
defined. In order to characterize both the average crack growth and
the stepwise growth dynamics, a statistical analysis is required. In
this section we  focus on the average dynamics (see also
ref.\cite{Santucci3}). The intermittent dynamics and in particular
the step size statistics will be discussed in the next section (see
also ref.\cite{Santucci2}).

\subsubsection{Statistically averaged crack growth}
We have performed an extensive study of crack growth, varying the
initial crack length from $L_i=1 \rm{cm}$ to $L_i=4\rm{cm}$ and the
applied force between $F=140 \rm{N}$ and $F=280\rm{N}$
(corresponding to an initial stress intensity factor between
$K_i=2.7\rm{MPa.m^{1/2}}$ and $K_i=4.2\rm{MPa.m^{1/2}}$) and
repeating $5$ to $20$ experiments in the same conditions (stress,
initial crack length, temperature and humidity rate). The resulting
measured lifetime varied from a few seconds to a few days depending
on the value of the applied stress or the temperature.
\begin{figure}[hbt!]
\centerline{{\resizebox{0.6\columnwidth}{!}{\includegraphics{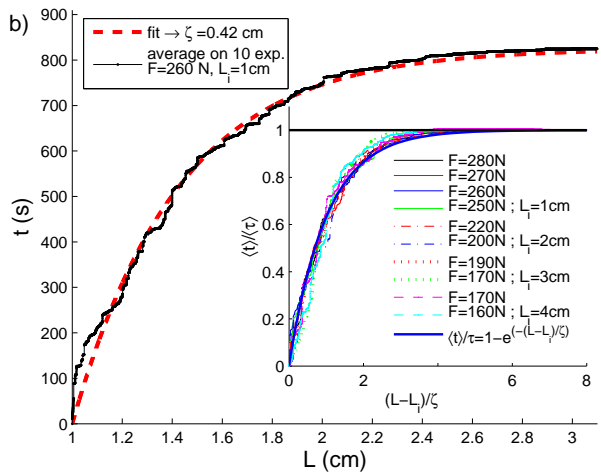}}}}
\caption{ Statistical average of growth curves for $10$ creep
experiments realized in the same conditions ($L_i=1\rm{cm}$,
$F=270\rm{N}$). The dashed line corresponds to a fit using equation
eq.~(\ref{eqgrowth}) with a single free parameter $\zeta = 0.41
\rm{cm}$. Insert: rescaled average time $\langle t \rangle/\tau$ as
function of rescaled crack length $(L-L_i)/\zeta$ for various
initial crack lengths and applied stress. The solid line corresponds
to eq.~(\ref{eqgrowth}).} \label{averagegrowth}
\end{figure}

In order to characterize the average growth dynamics, we examine for
given experimental conditions the average time $\langle t(L) \rangle
$ the crack takes to reach a length $L$, where $<.>$ stands for
ensemble average of $t(L)$ over many experiments. Even though the
lifetime distribution is large and the growth dynamics intermittent,
the average growth offers a regular behavior,  very close to the
exponential evolution given in eq.(\ref{eqgrowth}). Indeed, we
obtain a very good fit of the data in fig.~\ref{averagegrowth} with
eq.~(\ref{eqgrowth}) setting the mean lifetime to the experimentally
measured value and using $\zeta$ as a unique free parameter. We note
that the agreement is already quite good after averaging only $10$
experiments in the same conditions.

Using the same procedure, we extract the characteristic growth
length $\zeta$ for various experimental conditions. In the insert of
fig.~\ref{averagegrowth}, rescaling the crack length by $\zeta$ and
the time by $\tau$ for many different experimental conditions, we
show that the data collapse on the functional form given by
eq.~(\ref{eqgrowth}). Moreover, we have checked that the deviation
from the predicted average behavior reduces when increasing the
number of experiments.

\begin{figure}[hbt!]
\centerline{{\resizebox{0.5\columnwidth}{!}{\includegraphics{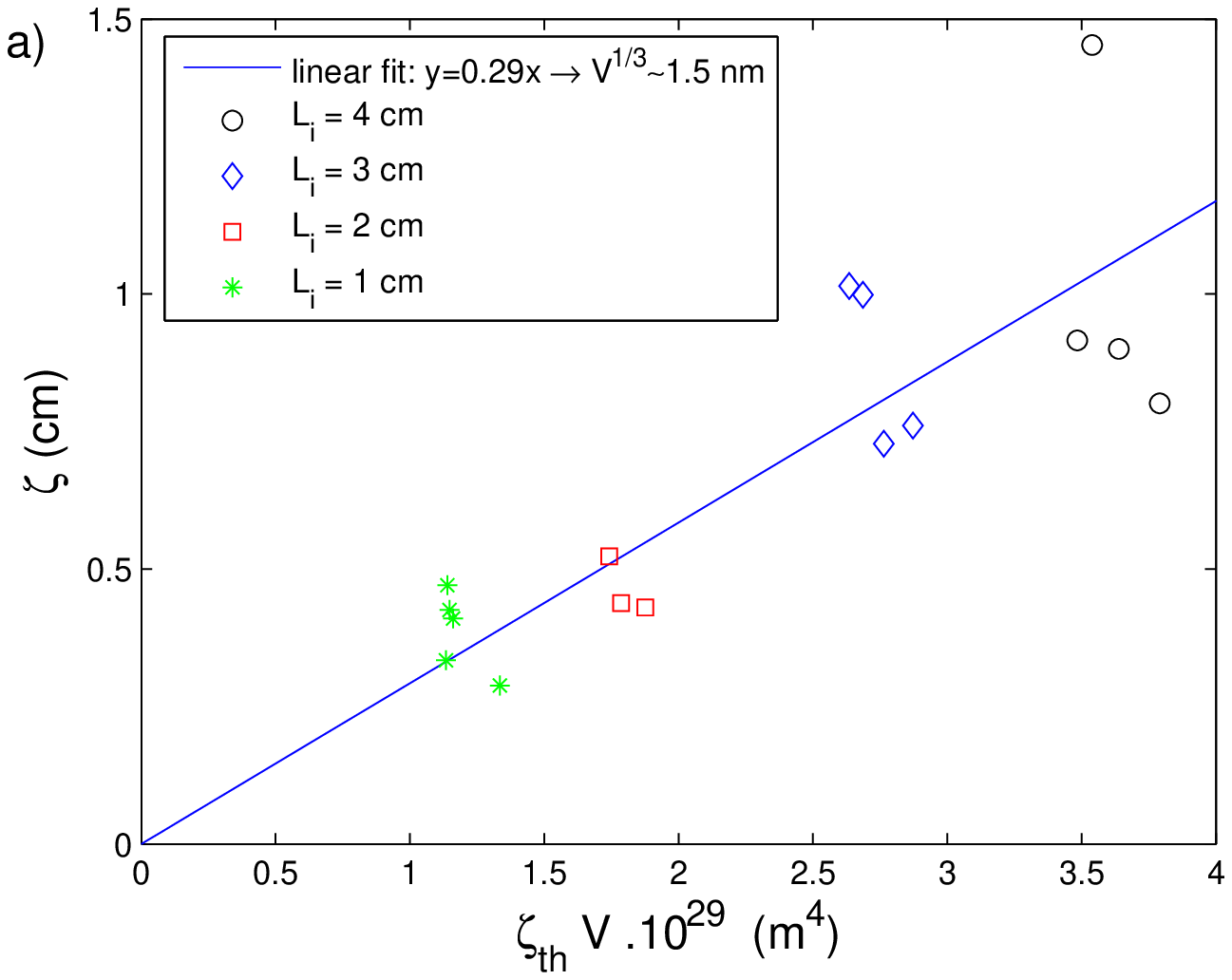}}}
{\resizebox{0.5\columnwidth}{!}{\includegraphics{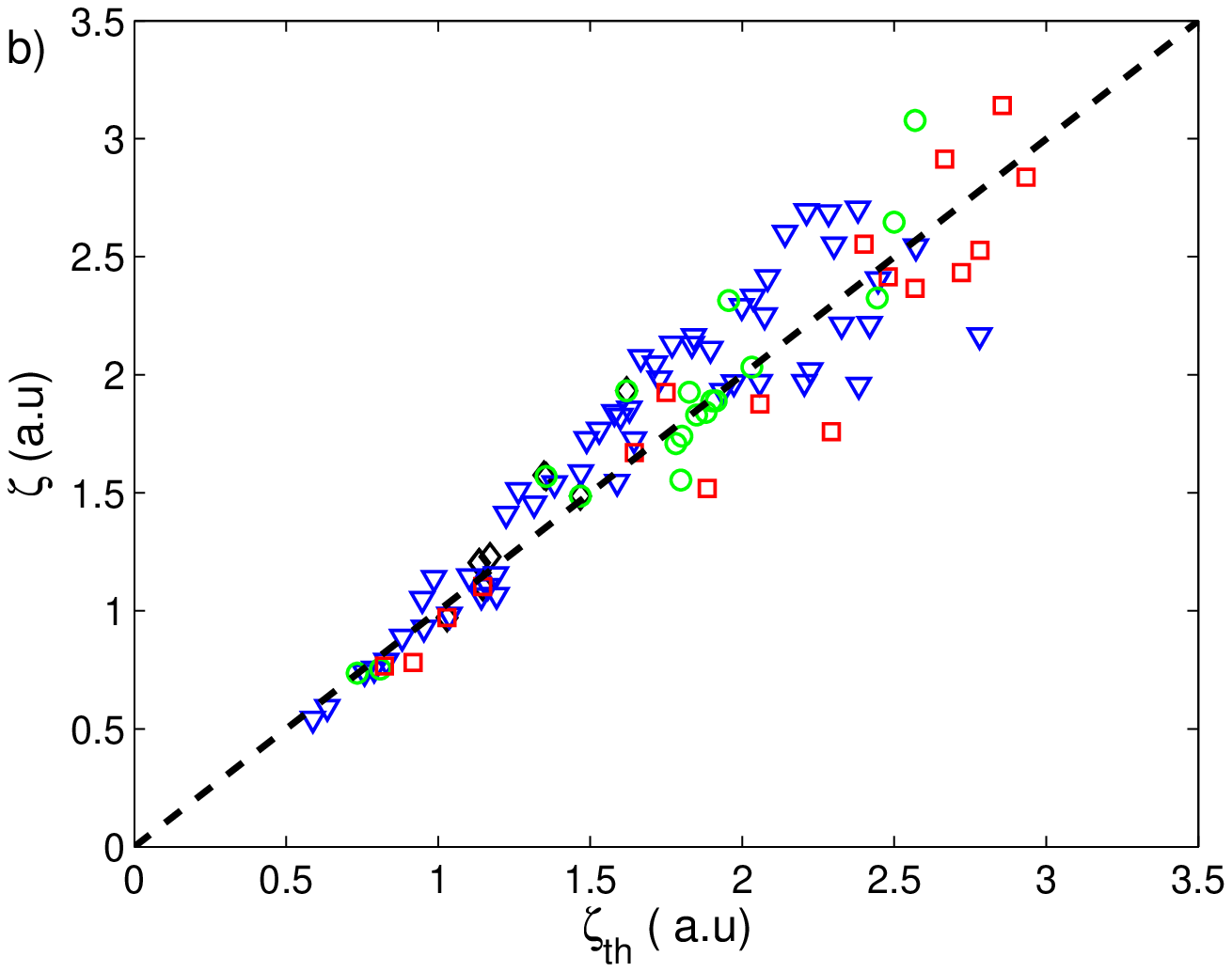}}}}
\caption{ a) Experimental value of $\zeta$ extracted from the
average growth profile as a function of the prediction of the
thermally activated rupture model. The line represents the best
linear fit $y=x/V$. Its slope permits us to obtain a characteristic
length scale for rupture: $V^{1/3}\sim 1.5\,\rm{nm}$. b) $\zeta$
extracted from the average growth profile from numerical simulations
\cite{Santucci1} versus the predictions of our model of activated
rupture. Each point corresponds to an average over $30$ numerical
experiments, at least. The solid line shows the behavior expected
from the model (slope $=1$). } \label{zeta}
\end{figure}

We now compare the measured $\zeta$ and $\tau$ with the theoretical
predictions of paragraph \ref{section_model}. By setting the value
of $\lambda$ to the maximum fiber diameter that is
$\lambda=50\mu\rm{m}$, we impose that there is no divergence of the
stress in the fiber and that the maximum stress is controlled by the
fiber size. Once $\lambda$ is fixed, the only adjustable parameter
is the activation volume $V$. In fig.~\ref{zeta}a) we plot the
measured values of $\zeta$ as a function of  the theoretical one
$\zeta_{th} V = 2 Y \lambda k_BT L_i /[K_i(K_c-K_i)]$ (see
eq.\ref{eqzeta}). Using $V$ as a free parameter, we find a
characteristic scale $V^{1/3}=1.5nm$ close to the microfibril
diameter $d$. On fig.~\ref{zeta}b, we show the value of $\zeta$
obtained by simulating the crack growth in a 2D square lattice of
springs (see \cite{Santucci1} for more details) and after averaging
over more than $30$ realizations of thermal noise. The agreement
with the analytical model is very good in average, but there is
still some dispersion due to the lack of statistics. We believe the
same lack of statistics affects the experimental results in
fig.~\ref{zeta}a. Interestingly, the dispersion seems to increase
with $\zeta$ both in the experiments and in the numerical
simulation.

For a fixed value of applied force $F$ and initial length $L_i$,
varying temperature between $20^{\circ}\rm{C}$ and
$120^{\circ}\rm{C}$ (symbols without error bar on fig.~\ref{zeta}b)
leads to variations of the rupture time up to four order of
magnitude. In fig.~\ref{zeta}b, we plot the rupture time as a
function of $\Delta U / V k_B T = (K_c-K_i)^2/(2 Y \ \lambda\ k_B
T)$ as predicted by our model (see eqs.\ref{eqtau},\ref{eqDU}). The
temperature that enters in this relation is the actual themodynamic
temperature, and not an effective temperature as in previous reports
\cite{Guarino99}. The error bars correspond to the experimental
dispersion of measured rupture times. We see that there is a rather
good collapse of the data whatever is the initial crack length
$L_i$. Fixing the value of $\lambda$, from a fit of the data, we
obtain independently a new estimate of $V$ which gives a
characteristic scale $V^{1/3}=2.2 nm$. Once again this estimate is
close to the microfibril diameter $d$.

\begin{figure}[hbt!]
\centerline{
{\resizebox{0.5\columnwidth}{!}{\includegraphics{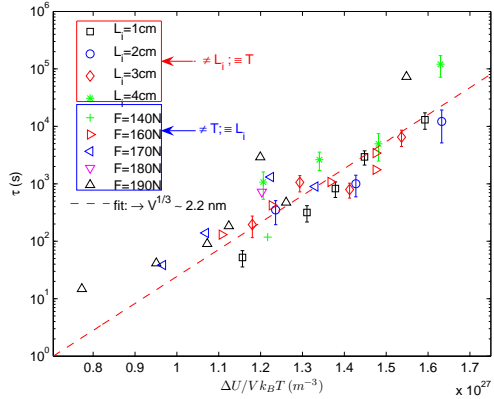}}}}
\caption{Logarithm of lifetimes as a function of the dimensional
factor $\Delta U/(V k_B T)$ predicted by eq.~\ref{eqtau} for
different values of $L_i$, $F$ and $T$. The slope of the best fit
$\log\,\tau\,\propto\,\Delta U / k_B T$ (dashed line) gives an
estimation of the characteristic length scale $V^{1/3} \sim 2.2
\rm{nm}$. Data points without error bars correspond to non-averaged
measurements obtained when varying temperature with $L_i=2cm$ and
various fixed values of $F$. } \label{fig.Lifetime}
\end{figure}

As a side remark, if we had considered rupture events were
reversible, we could have as well considered that the energy barrier
to overcome is given by the variation in Griffith potential energy
between the length $L_i$ and $L_G$: $\Delta E_G =
E_G(L_G)-E_G(L_i)$. Given the material parameters of paper and the
experimental conditions in which we observe rupture, this would have
given us a typical value $\Delta E_G/k_B T\sim 10^{18}$ and rupture
time virtually infinite. This enormous value, obviously physically
wrong, comes from ignoring the irreversible character of rupture
events.

\subsubsection{Comparison between the Griffith length $L_G$ and the critical length of rupture $L_C$}
We have observed on fig.\ref{singlegrowth}b) that the critical
length $L_c$ for which the paper sheet breaks suddenly, scales with
the inverse square of the applied stress $1/\sigma^2$. Thus, $L_c$
scales with $\sigma$ as the Griffith length $L_G=4Y\gamma / (\pi \
\sigma^2)$ does. In fact, we would normally expect that they are the
same length \cite{Griffith}. However, the lattice trapping model
presented in paragraph \ref{jump} predicts that $L_c$ and $L_G$ can
be two different lengths. In order to clarify which conclusion is
correct, we have designed a method to compute the Griffith length in
our experiments. For that purpose, we need to estimate the surface
energy $\gamma$ needed to open a crack.
\begin{figure}[hbt!]
\centerline{{\resizebox{0.5\columnwidth}{!}{\includegraphics{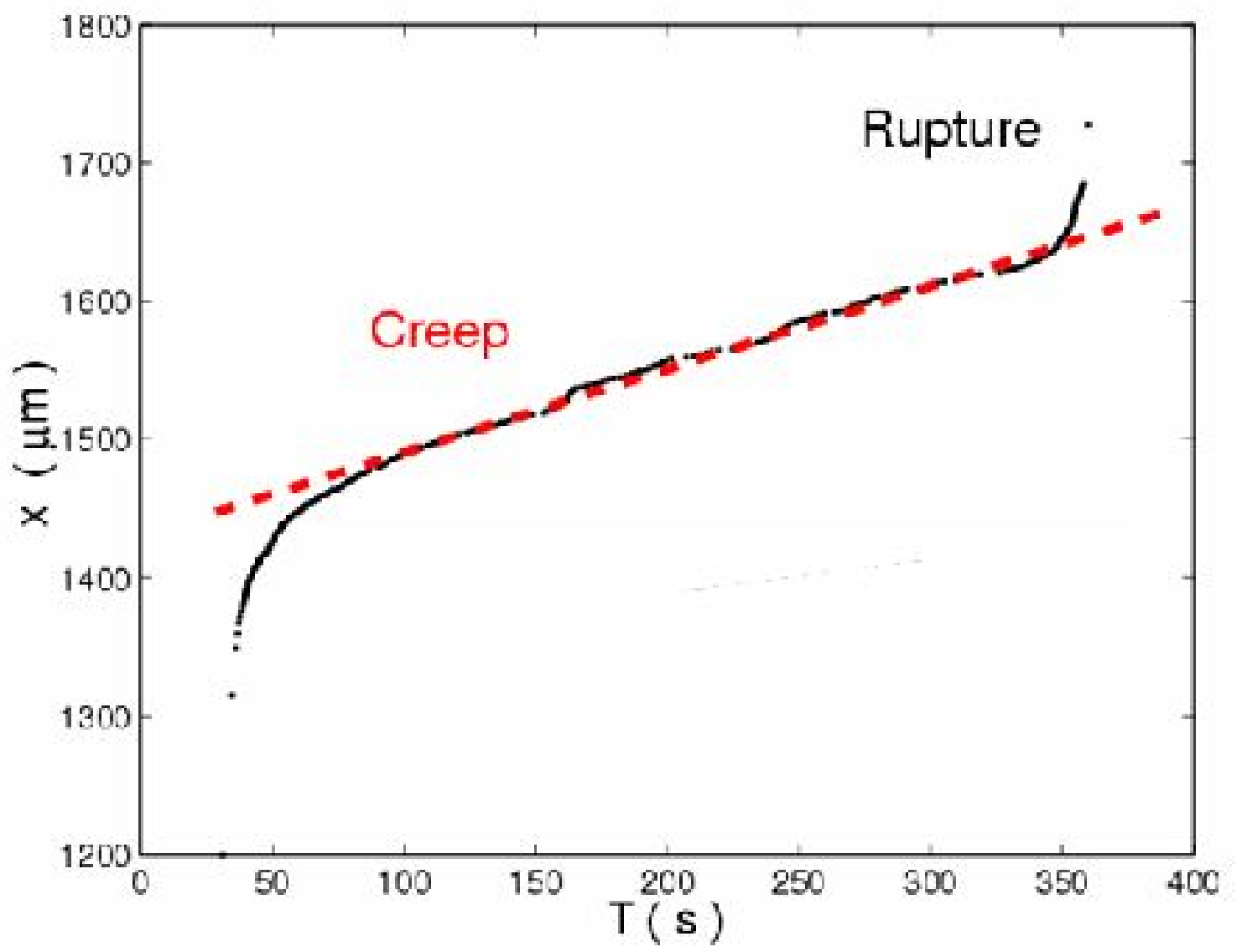}}}
{\resizebox{0.5\columnwidth}{!}{\includegraphics{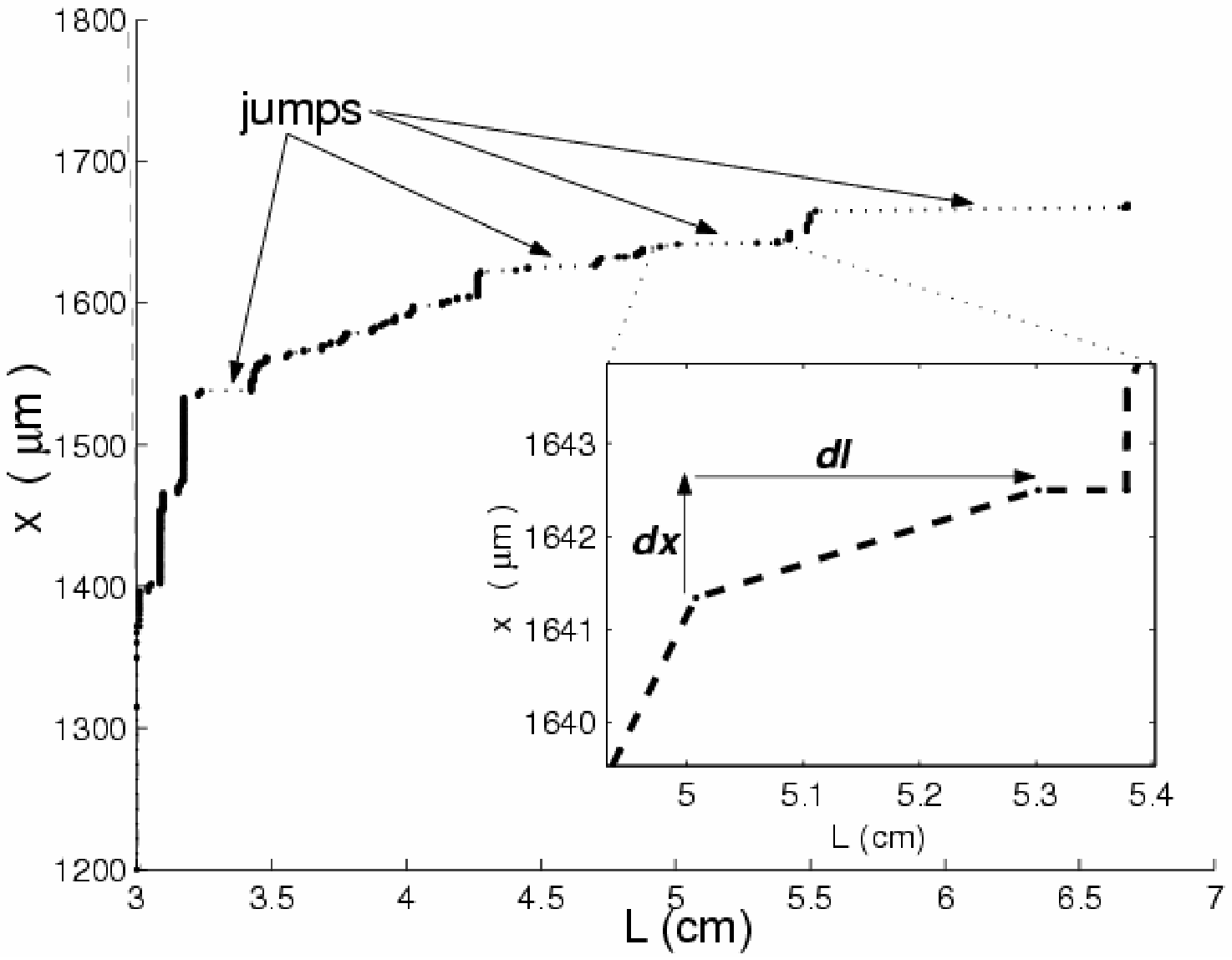}}}}
\caption{ a) Deformation of a sheet of paper with an initial defect
of $L_i = 3 \rm{cm}$ during an experiment at a constant load $F=190
\rm{N}$. b) Deformation of the sheet of paper as function of the
crack length for the same experiment.} \label{creep}
\end{figure}
Assuming that the work developed by the tensile machine will only
permit the crack to advance (we neglect any source of dissipation),
we can estimate an upper bound for the surface energy $\gamma$.
Indeed the minimum work provided by the tensile machine when making
a displacement $dx$ at a constant load $F$ in order to open the
crack of a length $dl$ is $\delta W_{min} = F dx = 4 \gamma e dl$.
Therefore we can give an upper limit for the surface energy :
$\gamma = F dx/4e dl$, when the crack advances of $dl$ for a
displacement $dx$ of the tensile machine, where $e$ is the width of
the sheet submitted to a constant force $F$.

However, during a creep experiment, we observe a slow global
deformation of the sheet of paper (see fig.\ref{creep}a) which is
uncorrelated with the crack growth. In order to suppress this creep
effect that will lead to an over-estimation of the surface energy
$\gamma$ we will examine the displacement of the tensile machine
$dx$ when the crack advances by jumps (see fig.\ref{creep}b)) which
occurs over much faster temporal scales than the creep deformation.
We can repeat this analysis for the various jumps detected during
the slow crack growth and finally for the various experiments
performed. Therefore, we obtain an average value for the surface
energy $\langle \gamma \rangle = 1540 \pm 180 \rm{N/m}$ (Note that
the dispersion of this measurement is really important, $\langle
\gamma^2 \rangle^{1/2}= 850 \rm{N/m}$).
\begin{figure}[hbt!]
\centerline{{\resizebox{0.38\columnwidth}{!}{\includegraphics{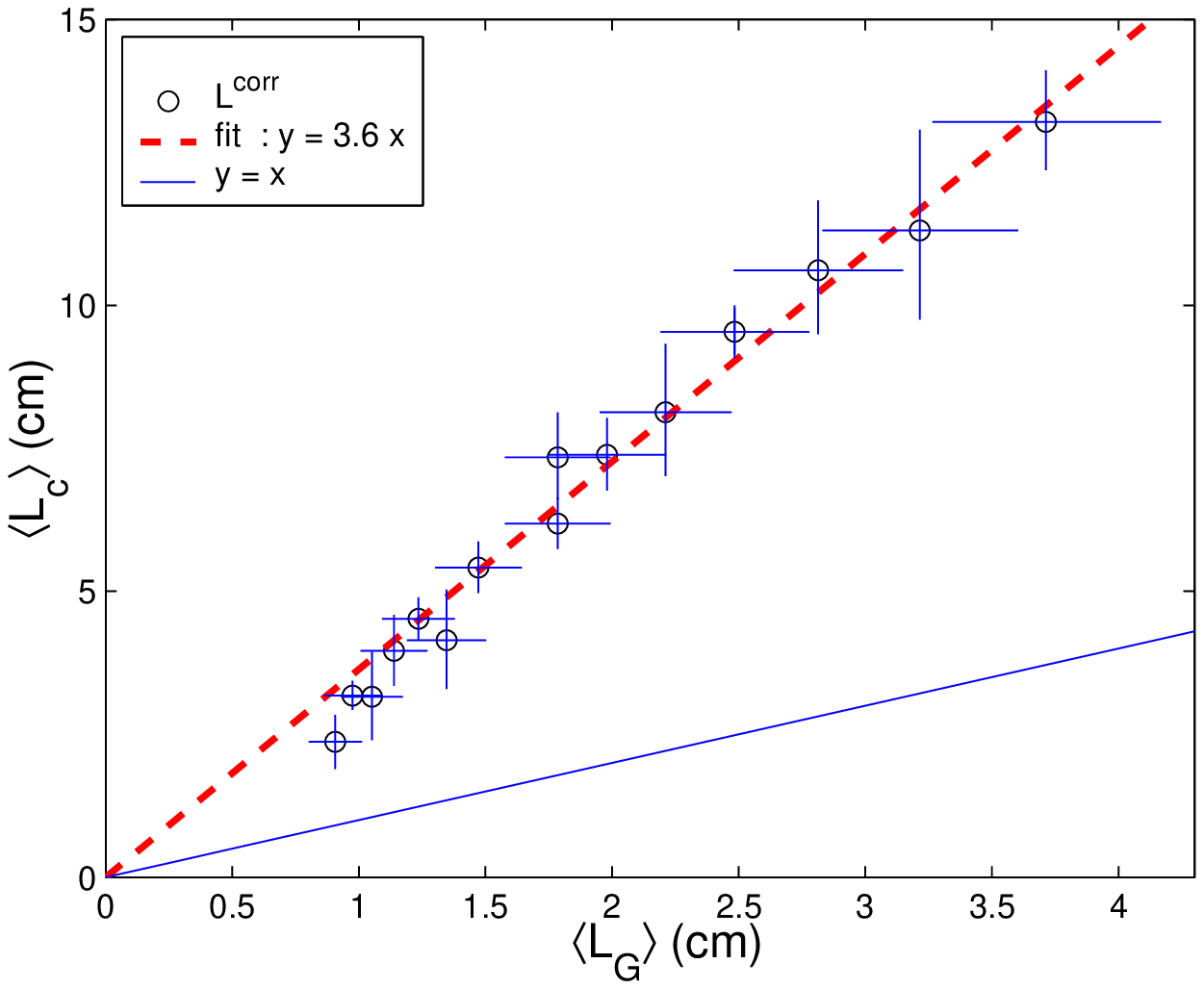}}}
{\resizebox{0.38\columnwidth}{!}{\includegraphics{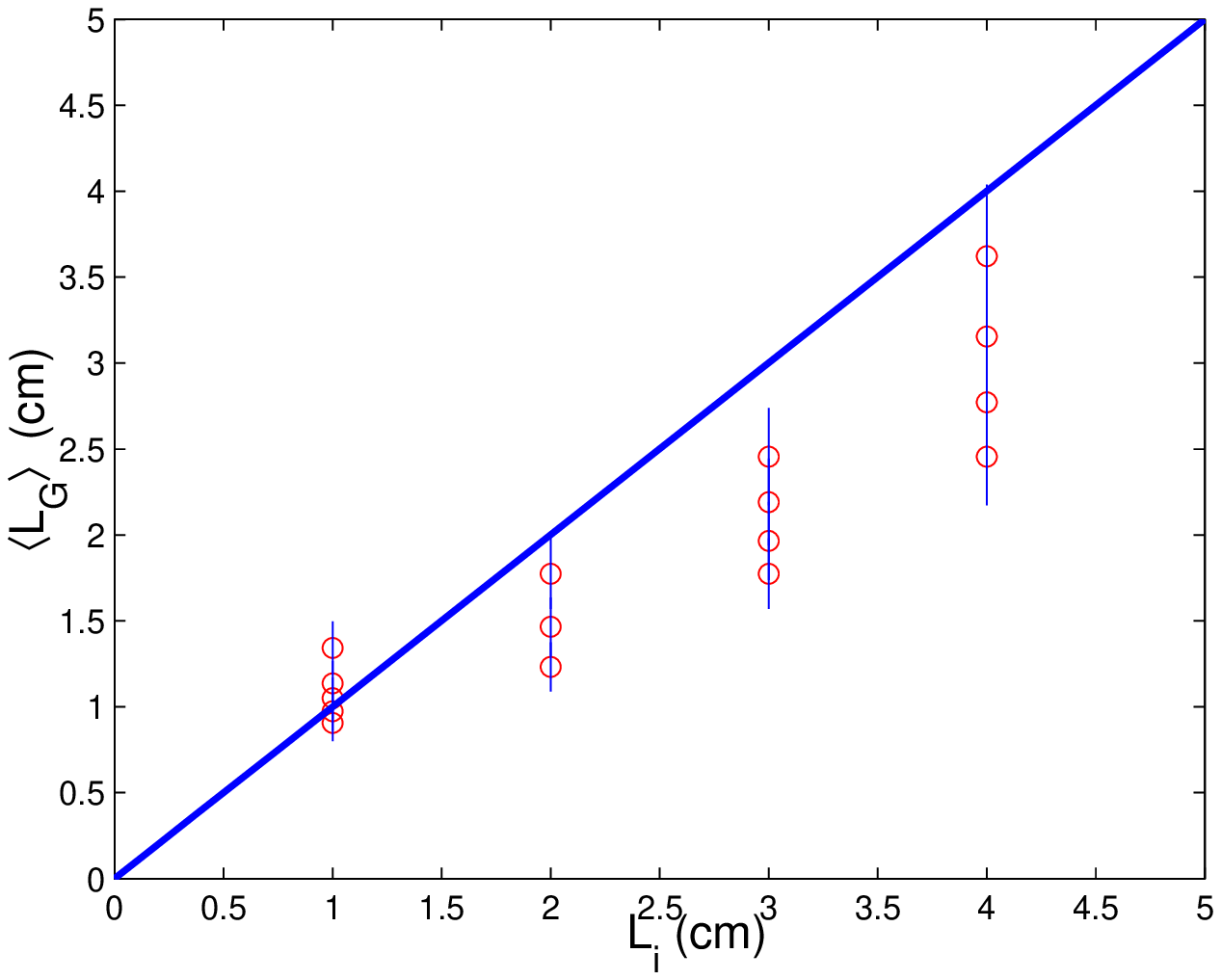}}}}
\caption{ a) Average critical length  $\langle L_c \rangle$ as
function of the Griffith $\langle L_G \rangle$. The dashed line
represents the best linear fit $y=3.6 x$ while the continuous line
is a guide for the eye showing $y=x$. b) Average Griffith length
$\langle L_G \rangle$ as function of the initial crack length $L_i$.
The line $y=x$ is a guide for the eye.} \label{LCLG}
\end{figure}

With this measurement of the surface energy we get an estimate of
the average Griffith length $\langle L_G \rangle$, using the Young
modulus of the fax paper sheets $Y = 3.3\,10^9 \rm{N.m^{-2}}$, and
finally we compare this length to the critical length measured
during our creep experiments where the paper sheet breaks suddenly.
Fig.\ref{LCLG}a) shows the average critical length $\langle L_c
\rangle $ as a function of the average Griffith length $\langle L_G
\rangle $ for the various creep experiments performed. We observe
clearly that the critical length for rupture is at least 3 times
larger that the critical length predicted by Griffith. Moreover it
is important to recall that we actually estimated an upper limit of
this Griffith length since the slow deformation of the sample during
the creep experiments leads to an over-estimation of the surface
energy needed to open a crack which was also by definition an upper
bound of the surface energy of our samples. Thus, we could expect
the Griffith length to be even smaller. We can also notice on
fig.\ref{LCLG}b) that the Griffith length is for all the various
experiments performed smaller or very close to the initial crack
length $L_i$. This suggests that $L_G$ might be indeed a critical
length below which no crack propagation can occur, at least during a
reasonable experimental time.

\subsubsection{Comparison with the model describing the statistically averaged crack growth}
The experimental measurements of the mean crack growth in paper
sheets show a rather good agreement with the predictions of the
activation model described in \ref{section_model}. Indeed the model
predicts the evolution of the crack length as a function of time,
the characteristic length $\zeta$ and the life time $\tau$ of the
sample as a function of the applied stress and the temperature. It
is important to notice that the only free adjustable parameter is
the activation volume $V$ which represents the scale at which the
statistical stress fluctuations trigger rupture events. This
activation volume $V$ turns out to be about the same, within error
bars, in the fits of two very different measured physical quantities
$\tau$ and $\zeta$ as a function of the theoretical predictions. It
is an interesting experimental evidence that $V^{1/3}$ is extremely
close to the diameter $d$ of the microfibrils that make up the
macroscopic paper fibers \cite{Jakob}, suggesting that the rupture
first occurs inside a fiber at the nanometric scale of the
microfibrils and then progressively leads to the rupture of the
paper fiber itself. It is possible to include such a progressive
rupture mechanism of the macroscopic fiber in the rupture model that
we discuss in this paper. Indeed, the rupture time of a bundle of
micofibrils is dominated by the same exponential factor than the one
used for $\tau_V$ in equation (\ref{motion}). It was argued in
\cite{Santucci06} that the volume at which rupture occurs is then
$V=d^3$ while the stress level in the microfibrils is set by the
whole bundle diameter $\lambda \simeq \sqrt{n}d$, where $n$ is the
number of microfibrils making up the macroscopic fiber. It is rather
commonly observed that the subcritical time of rupture is controlled
by thermally activated mechanisms at the nanometer scale
\cite{Zhurkov,Bueche1957}. In addition to connecting this nanometer
scale to a microstructure of the material, we show that it is also
possible to describe the whole growth dynamics in a brittle material
using the ambient temperature for $T$, without having to take into
account an effect of disorder through an effective temperature
\cite{Scorretti}.

Moreover, we show that, in our creep experiments, the Griffith
critical length $L_G$ corresponding to the maximum of the Griffith
potential energy is smaller than the critical length of rupture
$L_C$ as well as the initial crack length $L_i$. This result is in
agreement with previous predictions \cite{Marder}, and with our
numerical simulations on a 2D elastic spring network. It also
appears consistent with the picture described in section \ref{jump}
where, due to the lattice trapping, a crack with a length above the
Griffith length will grow irreversibly even if one assumes that the
rupture is reversible.

\section{Statistics of the jumps of the crack tip}
Looking at the typical growth curve plotted in
Fig.\ref{singlegrowth}a) it clearly appears that the crack does not
grow smoothly: essentially, there are periods of rest where the
crack tip does not move and periods where it suddenly opens and
advances of a certain step size $s$. In
sect.\ref{section_experiment} we have seen that varying the initial
crack length ($1cm<L_i<4cm$) and the initial stress intensity factor
$K_i$ between $2.7 \rm{MPa.m}^{1/2}$ and $4.2\rm{MPa.m}^{1/2}$, the
resulting measured lifetime varied from a few seconds to a few days
depending on the value of the applied stress or the temperature.
Even for the same experimental conditions (same stress, initial
crack length and temperature) a strong dispersion in lifetime is
observed. This is of course expected for a model of thermally
activated growth \cite{Santucci1} as the one discussed in
sect.\ref{section_model} which  describes the mean behavior of the
crack, as we have shown in sect.\ref{section_experiment}. Here, we
want to study more extensively the step size statistics and to check
if our model may describe the distributions of the jump amplitudes.

\subsection{Experimental results}
In sect.\ref{section_experiment} we have seen that the crack
velocity is an increasing function of the stress intensity factor
$K$. Thus, it is natural to look at the step statistics for a given
value of $K$. We have seen  also that the $p(s)$ predicted by our
model (eq.\ref{distri}) is a function of $K$. In practice, the step
size distributions have been obtained for various ranges of $K$.

\begin{figure}[hbt!]
\centerline{{\resizebox{0.5\columnwidth}{!}{\includegraphics{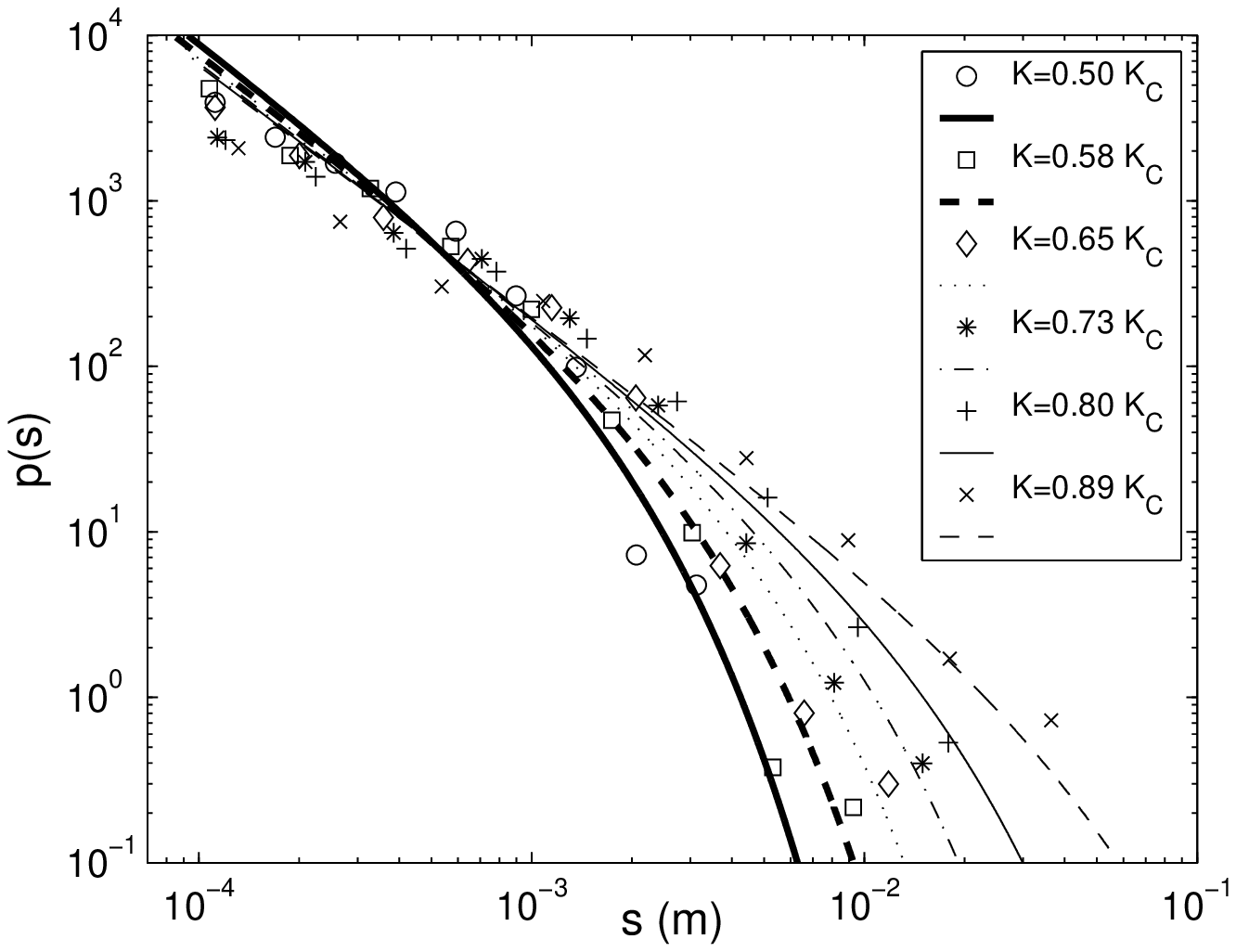}}}
{\resizebox{0.45\columnwidth}{!}{\includegraphics{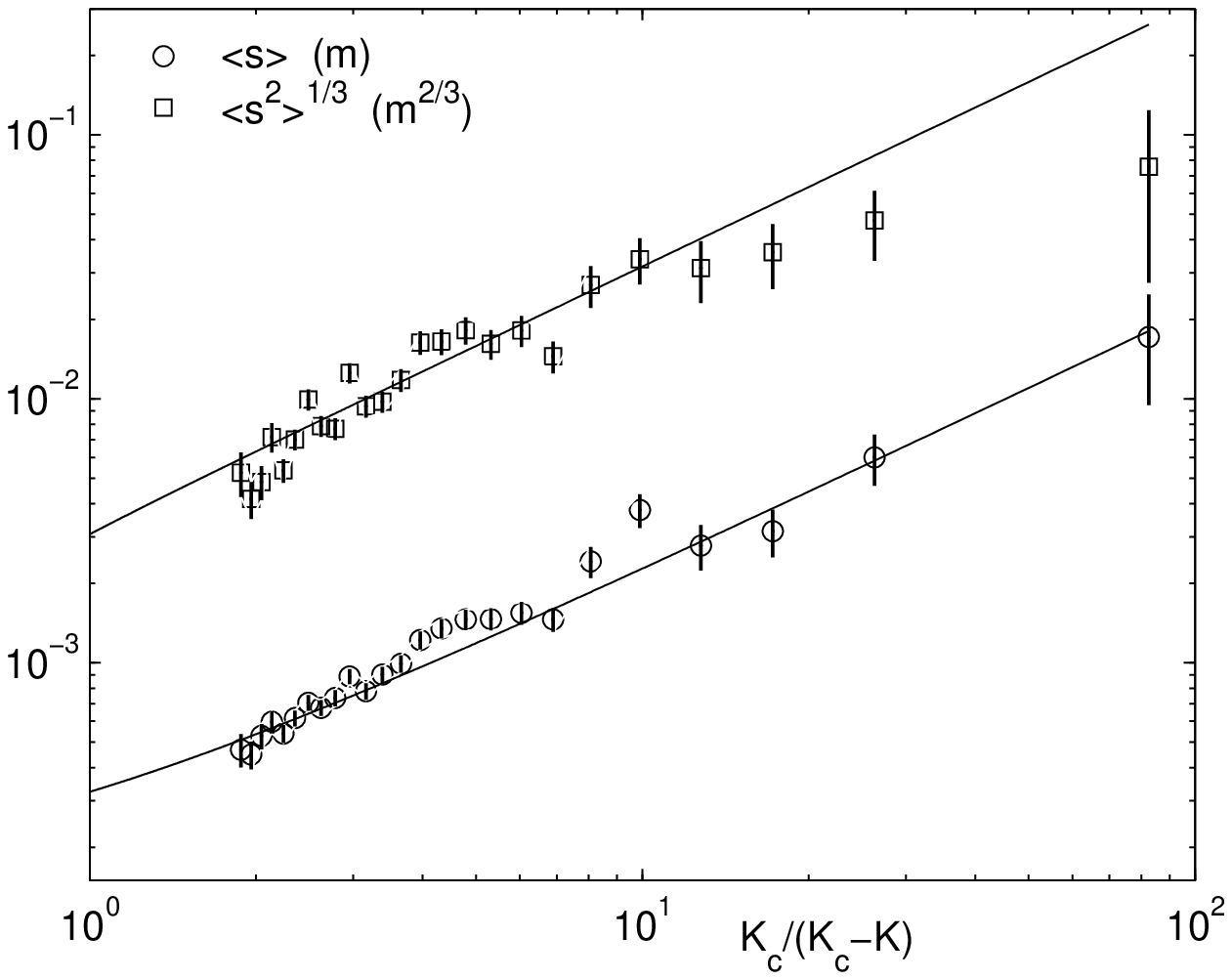}}}}
\caption{a) Probability distribution of step sizes for various
values of stress intensity factor. Choosing $\lambda=50 \mu \rm{m}$,
the different curves are the best fits of eq.\ref{distri} giving an
average value $V=(5\pm 1 \rm{ })$ {\AA}$^3$. b)The mean and cubic
root of the variance from raw measurements of step sizes is well
reproduced by the model (eq.(\ref{mean}) and eq.(\ref{var})) plotted
with $\lambda=50\mu \rm{m}$ and $V=5$ \AA$^3$.}
\label{steps_moments}
\end{figure}

Fig.\ref{steps_moments}a) shows the step size distributions
determined from all the data we have collected using a logarithmic
binning. Typically, $700$ data points are used to obtain each
distribution. Two regimes are observed. For small step sizes, the
distribution does not depend on the value of K, while for larger
step sizes there is a cut-off size increasing with $K$. In practice,
the toughness of the material, i.e. its critical stress intensity
factor $K_c=6.5\pm 0.05\rm{MPa.m}^{1/2}$, has been obtained as the
value of $K$ beyond which the probability to detect a jump vanishes.

The normalization condition of the distribution actually reduces the
model to one parameter, the ratio $V/\lambda^2$. As we have already
done in sect.\ref{section_experiment}
 we fix $\lambda=50\mu\rm{m}$ and  $V$ is the only unknown. Using for $p(s)$
 the expression derived in eq.\ref{distri}, one parameter fits
of step size distributions in Fig.\ref{steps_moments}a) for each
range of stress intensity factors give very robust results. We find
$V=5\pm 1$ {\AA}$^3$ which is quite different from  the value
obtained from the fits of the average dynamics in
sect.\ref{section_experiment}. The possible reasons of this
difference will be discussed in the next section. Here  we focus on
the comparison between the computed $p(s)$ eq.\ref{distri} and the
measured one. To check the asymptotic limit close to the critical
point, we have plotted in Fig.\ref{steps_moments} $\langle s
\rangle$ and $\langle s^2 \rangle^{1/3}$ as a function of
$K_c/(K_c-K_m)$. Here, the mean and the variance of step sizes have
been computed from the raw measurements in a given range of $K$.
Because it requires less statistics to estimate the first two
moments of the distribution than the distribution itself, we are
able to narrow the width of the $K$ range for each data point
without changing the global trend. The solid lines represents the
model prediction using the fitted value of $V$ from the
distributions of Fig.\ref{steps_moments}. Not only the model
reproduces reasonably well the evolution of the step size
distributions with $K$ ($V$ is essentially constant and all the
other parameters are fixed), but the asymptotic divergence of the
first two moments of the distribution are also well reproduced. For
the mean step size, the scaling is observed up to $K$ values very
close to $K_c$, about $1\%$. In the model, we see that the ratio of
the standard deviation of the distribution over the mean size is
diverging at $K_c$. Thus, close to $K_c$, the measure of variance
becomes more inaccurate than for the mean.

\subsection{Limits of this approach}
\begin{figure}[hbt!]
\centerline{{\resizebox{0.45\columnwidth}{!}{\includegraphics{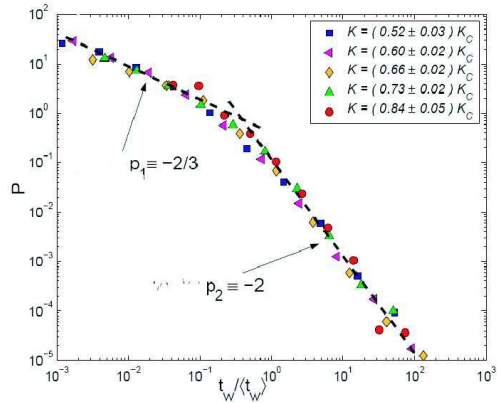}}}}
\caption{Distribution of waiting times between two crack jumps. All
the distributions for various ranges of stress intensity factor
collapse on a single curve when normalized by the average waiting
time value corresponding to each range.}\label{waiting}
\end{figure}

We note that some experimental observations are not taken into
account by our model. For instance, it could have been expected from
\cite{Roux} that the distribution of waiting times should be
exponential. On Fig.\ref{waiting}, we see that the distribution is
not exponential. As a guide for the eye, we show that for small
times the slope in log-log scale is around $2/3$ and for larger
times $2$. Also, the model does not consider the roughness that the
crack develops and the effect it has on the local stress field. This
will change the dynamics of the crack growth and it would be
interesting to make a connection with the statistical properties of
the crack roughness discussed in
\cite{Bouchbinder,Santucci06,Mallick}.

More important, we have noticed in the previous section that the
value of $V$ obtained from the fit of the measured $p(s)$ is at the
atomic scale, i.e. $V^{1/3}\simeq 1.7$ \AA. This scale is one order
of magnitude below the one obtained from the fits of the average
dynamics in sect.\ref{section_experiment}. There are several reasons
for this difference. It should be realized that the model for the
jumps actually predicts a lower limit for this microscopic scale.
Indeed $s\propto U_f \lambda/U_c$ is a simple dimensional argument
and the prefactor is unknown. We arbitrarily fixed it to 1, but this
prefactor could be much larger. Our choice implicitly assumes that
there is a strong dissipation of energy during crack advance since
none of the elastic release of energy is used to keep the crack
moving,  This is certainly an overestimation of a real dissipative
mechanism, would it be viscoelastic or plastic. Decreasing
dissipation in the model will permit larger steps of the crack. In
order to obtain the same experimental velocity, the trapping time
must also be larger which will happen if the rupture occurs at a
larger microscopic scale. The other point that has been neglected is
the disorder in the material properties. It has been shown recently
that disorder effectively reduce the energy cost for breaking and
this will also permit rupture at a larger microscopic scale
\cite{Scorretti}. However, several recent works have shown that when
a macroscopic crack is growing, the disorder will actually help pin
the crack and slows down its dynamic \cite{Cortet06,Kierfeld06}.
Thus, further theoretical work needs to be done mainly with the aim
of introducing a more realistic dissipative mechanism.

Despite these limitations, we believe that our approach in the
present form is a first step to describe rupture in brittle
materials for which a structure at a mesoscopic scale exists. At
this point, the rupture model presented in this paper can not
describe rupture in ductile materials and we expect that taking into
account plasticity, for instance, will lead to significantly
different growth dynamics \cite{Cortet05}.

\section{Conclusion}

To conclude, we have seen that the slow crack dynamics in fibrous
materials such as paper is a complex and rich statistical process.
Specifically, we have studied the sub-critical growth of a single
crack during creep experiments and observed stepwise growth
dynamics. Despite this complexity, a statistical average of the
growth dynamics reveals a simple behavior. We have shown that a
simple model of irreversible and thermally activated rupture is able
to predict with good accuracy the average crack growth observed,
which can be characterized by only two parameters: the rupture time
$\tau$ and a characteristic growth length $\zeta$. Both quantities
are in reasonable agreement with the model. In particular, we
verified experimentally that rupture time depends exponentially on
temperature as previously observed \cite{Zhurkov}. The comparison of
our experimental results on the averaged crack growth to the
thermally activated rupture model suggests that the thermodynamical
stress fluctuations have a proper amplitude to trigger rupture
events at a nanometric scale corresponding to the width of cellulose
microfibrils. This is consistently found from two independent
measurements : the rupture time $\tau$ and a characteristic growth
length $\zeta$.

Moreover, we have shown \cite{Santucci2} that we can adapt and
extend our first model \cite{Santucci1} in order to describe the
stepwise growth dynamics in a rugged potential energy landscape. The
microstructure of our samples could indeed modify the Griffith
energy barrier leading to a lattice trapping effect
\cite{Thomson86}). As a consequence this modified model predicts the
step size statistics. This is quite interesting because it may open
new perspectives in the description of rupture as a thermally
activated process. However concerning this last approach more work
is needed to introduce a more realistic dissipative mechanism in the
material properties in order to understand the reasons why the
activation volume obtained from the jump statistics (section 4) is
smaller than the volume obtained obtained from the average crack
growth properties in (section 3).

\vskip 1cm

{\bf Acknowledgements} This paper is dedicated to the memory of
Carlos Perez Garcia. One of us (S.C.) remembers him not only for the
very interesting scientific collaborations and useful discussions
but mainly for his kindness and warm friendship.


\begin{thebibliography}{}
\bibitem{herrmann90}
 H. J. Herrmann, S. Roux, {\sl Statistical models for the fracture of
disordered media\/} (Elsevier, Amsterdam, 1990).

\bibitem{Alava}
M. J. Alava, P. K. N. N. Nukala, S. Zapperi, Adv. in Phys.
\textbf{55}, 349-476 (2006).

\bibitem{Brenner} S. S. Brenner, J. Appl. Phys. \textbf{33}, 33 (1962).

\bibitem{Zhurkov} S. N. Zhurkov, Int. J. Fract. Mech. \textbf{1}, 311 (1965).

\bibitem{Golubovic} L. Golubovic, S. Feng, Phys. Rev. A \textbf{43},
5223 (1991).

\bibitem{Pomeau1} Y. Pomeau, C.R. Acad. Sci. Paris II \textbf{314} 553 (1992); C.R.
M\'ecanique \textbf{330}, 1 (2002).

\bibitem{Buchel} A. Buchel, J. P. Sethna, Phys. Rev. Lett. \textbf{77}, 1520 (1996); Phys. Rev. E \textbf{55}, 7669 (1997).

\bibitem{Kitamura97} K. Kitamura I. L. Maksimov, K. Nishioka, Phil. Mag. Lett. \textbf{75}, 343 (1997).

\bibitem{Roux} S. Roux, Phys. Rev. E \textbf{62}, 6164 (2000).

\bibitem{Scorretti} R. Scorretti, S. Ciliberto, A. Guarino, Europhys. Lett. \textbf{55}(5), 626 (2001);

\bibitem{Santucci1} S. Santucci, L. Vanel, R. Scorretti, A. Guarino, S. Ciliberto, Europhys. Lett. \textbf{62}, 320
(2003).

\bibitem{Schapery86} R. A. Schapery, \textit{In Encyclopedia of Material Science and
Engineering} (Pergamon, Oxford, 1986), p. 5043.

\bibitem{Langer} J. S. Langer, Phys. Rev. Lett. \textbf{70}, 3592 (1993).

\bibitem{Chudnovsky} A. Chudnovsky, Y. Shulkin, Int. J. of Fract. \textbf{97}, 83 (1999).

\bibitem{Paris} P. Paris, F. Erdogan, J. Basic Eng., \textbf{89}, 528 (1963).

\bibitem{Griffith} A. A. Griffith, Phil. Trans. Roy. Soc. Lond. A \textbf{221}, 163 (1920).

\bibitem{Lawn} B. Lawn , T. Wilshaw, \textit{Fracture of Brittle Solids} (Cambridge University Press,
Cambridge, 1975).

\bibitem{Pauchard} L. Pauchard, J. Meunier, Phys. Rev. Lett. \textbf{70}, 3565 (1993).

\bibitem{Ciliberto} S. Ciliberto, A. Guarino, R. Scorretti, Physica D \textbf{158}, 83 (2001).

\bibitem{Diu}
B. Diu , C. Guthmann, D. Lederer, B. Roulet, \textit{Physique
Statistique} (Herrmann, Paris 1989), 272.

\bibitem{Marder} M. Marder, Phys. Rev. E \textbf{54}, 3442 (1996).

\bibitem{Thomson86} R. Thomson, in \textit{Solid State Physics}, edited by H. Ehrenreich and D. Turnbull (Academic, New York, 1986), Vol. 39, p. 1.

\bibitem{Stauffer91} D. Stauffer, \textit{Introduction to Percolation
Theory} (Taylor \& Francis, London, 1991).

\bibitem{Santucci3} S. Santucci, P-P. Cortet, S. Deschanel, L. Vanel, S.
Ciliberto, Europhys.  Lett. \textbf{74} (4), 595 (2006).

\bibitem{Santucci2}
S. Santucci, L. Vanel, S. Ciliberto, Phys. Rev. Lett. \textbf{93},
095505 (2004).

\bibitem{Guarino99} A. Guarino, S. Ciliberto, A. Garcimart\`{\i}n, Europhys. Lett. \textbf{47} (4), 456 (1999).
\bibitem{Jakob} H. F. Jakob, S. E. Tschegg, P. Fratzl, J. Struct. Biol. \textbf{113}, 13 (1994).

\bibitem{Cortet06} P.-P. Cortet, L. Vanel, S. Ciliberto, Europhys. Lett. \textbf{74}, 602 (2006).

\bibitem{Kierfeld06} J. Kierfeld, V. M. Vinokur, Phys. Rev.
Lett. \textbf{96}, 175502 (2006).


\bibitem{Bouchbinder} E. Bouchbinder, I. Procaccia, S. Santucci, L. Vanel, Phys. Rev. Lett. \textbf{96}, 055509
(2006).

\bibitem{Santucci06} S. Santucci, K. J. M{\aa}loy, A. Delaplace, J. Mathiesen, A. Hansen,
J. O. Haavig Bakke, J. Schmittbuhl, L. Vanel, P. Ray, Phys. Rev. E.,
\textbf{75} 016104 (2007).

\bibitem{Bueche1957} F. Bueche, J. App. Phys. \textbf{28}(7), 784
(1957).

\bibitem{Mallick} N. Mallick, P-P. Cortet, S. Santucci, S. Roux, L. Vanel.
 submitted. to Phys. Rev. Lett. (2006).

\bibitem{Cortet05}
P.P. Cortet, S. Santucci, L.Vanel, S. Ciliberto, Europhys. Lett.
\textbf{71} (2), 1 (2005).

\end{thebibliography}
\end{document}